\documentclass[aps,twocolumn,pra,amssymb,floatfix,superscriptaddress,10pt]{revtex4}

\usepackage{graphicx}
\usepackage{mathptmx}

\newcommand{\njtf}{{n}_{j {\rm (TF)}}}
\newcommand{\tnjtf}{\tilde{n}_{j {\rm (TF)}}}
\newcommand{\ninj}{\tilde{n}_{1 {\rm (inj)}}}

\begin{document}

\title{Quantum swapping of immiscible Bose-Einstein condensates as an alternative to the Rayleigh-Taylor instability}

\author{D. Kobyakov}
\affiliation{Department of Physics, Ume{\aa} University, 901 87 Ume{\aa}, Sweden}

\author{A. Bezett}
\affiliation{Institute for Theoretical Physics, Utrecht University, Leuvanlaan 4,3584 CE Utrecht, The Netherlands}

\author{E. Lundh}
\affiliation{Department of Physics, Ume{\aa} University, 901 87 Ume{\aa}, Sweden}

\author{M. Marklund}
\affiliation{Department of Physics, Ume{\aa} University, 901 87 Ume{\aa}, Sweden}

\author{V. Bychkov}
\affiliation{Department of Physics, Ume{\aa} University, 901 87 Ume{\aa}, Sweden}

\begin{abstract}
We consider a two-component Bose-Einstein condensate in a quasi-one-dimensional harmonic trap, where the immiscible components are pressed against each other by an external magnetic force. The zero-temperature non-stationary Gross-Pitaevskii equations are solved numerically; analytical models are developed for the key steps in the process. We demonstrate that if the magnetic force is strong enough, then the condensates may swap their places in the trap due to dynamic quantum interpenetration of the nonlinear matter waves. The swapping is accompanied by
development of a modulational instability leading to quasi-turbulent excitations. Unlike the multidimensional Rayleigh-Taylor instability in a similar geometry of two-component quantum fluid systems, quantum interpenetration has no classical analogue.
A crossover between the Rayleigh-Taylor instability and the quantum interpenetration in a two-dimensional geometry is demonstrated.
\end{abstract}

\maketitle

\section{Introduction} Bose-Einstein condensates (BECs) of dilute ultracold gases are ideal systems to study a wide range of many-body quantum phenomena both experimentally and theoretically.
Recently, there has been much interest in the quantum hydrodynamics of two-component BECs \cite{Papp-et-al-2008_Miscibility,Leslie-et-al-2009_fluctuations,Tojo-et-al-2010_separation,Takeuchi-et-el-2010_Counterflow,bib_Bubbles_in_BEC} and, in particular, in the development of  quasi-hydrodynamic instabilities at the interface separating the immiscible components
\cite{Blaaugeers-et-al-KH-exper,Saito-Ferrofluid-BEC,Sasaki-2009-RT,Bezett-et-al-RM,Takeuchi-2010-KH,bib_KH_BECs,Kobyakov-2011-linear}.
Among these phenomena, the quantum Rayleigh-Taylor instability (RTI) is, probably, one of the most representative and fascinating \cite{Sasaki-2009-RT,Kobyakov-2011-linear}. The RTI releases the excessive potential energy stored in the system and transforms it into kinetic energy of the flow, by means of interface waves that grow into mushroom-shaped bubbles.  In the classical case, the RTI develops when a light fluid (gas, plasma) supports a heavy one in a gravitational field, real or effective, as is the case, for example, in inertial confined fusion and Supernovae \cite{Kull,Cabot-Cook,Bychkov-et-al-RT-review,Modestov-et-al-2008}. In classical hydrodynamics, two immiscible fluids  with negligible diffusion (e.g., water and oil) may exchange places and reduce the potential energy only with the help of the RTI.

The quantum counterpart of the RTI was suggested by Sasaki et al. \cite{Sasaki-2009-RT} for a system of two phase-separated BECs with opposite projections of the hyperfine spin placed in a non-uniform external magnetic field, which  pushes the components against each other.
Recent results on the linear and nonlinear stages of the quantum RTI may be found in Refs. \cite{Sasaki-2009-RT,Kobyakov-2011-linear}.
 Still, contrary to the classical case, the RTI is not the only way for the immiscible BEC components to exchange places. As we show in the present paper, the transformation of potential energy into kinetic may also happen by way of quantum interpenetration of the immiscible condensates.
A similar effect was encountered recently in Ref. \cite{Gautam-Angom} in a simulation of a system consisting of a layer of ${}^{85}\text{Rb}$ sandwiched between two layers of ${}^{87}\text{Rb}$ in a 2D pancake-shaped trap.
Upon abrupt increase of the scattering length of the ${}^{85}\text{Rb}$ component, it expanded from the center to the outer parts of the trap. The resulting evolution of the density pattern  did not resemble the RTI at all. Although the process was presented as development of the RTI, we would like to point out that Ref. \cite{Gautam-Angom} actually demonstrated dynamical quantum interpenetration of the BEC components moving from the unstable to stable positions in an almost one-dimensional (1D) regime.

The purpose of the present paper is to  clarify the purely quantum nature of the encountered phenomenon. We demonstrate the possibility of quantum dynamical interpenetration of two immiscible BECs pressed against each other by means of an external potential in 1D geometry, which eliminates the intrinsically multidimensional RTI. We also demonstrate a crossover between the quantum dynamic interpenetration and the RTI in the two-component BEC system for a 2D case.

The paper is organized as follows. In Sec. \ \ref{sec:model} we formulate the basic equations and parameters describing the problem. In Sec.\ \ref{sec:onedim}, we present numerical results for the 1D case, accompanied by simplified analytical models to advance the understanding. In the process of doing this, we find an expression for the breathing mode frequency of two phase-separated BECs, a result that has so far been missing in the literature. In Sec.\ \ref{sec:twodim}, we reproduce the effect in a 2D geometry and compare the process of quantum interpenetration to the quantum RTI. Finally, in Sec.\ \ref{sec:summary}, we summarize and conclude.

\section{Basic equations describing a two-component BEC }
\label{sec:model}
We consider a harmonically trapped ferromagnetic spin-1 BEC with equal population of the $|1,\pm1\rangle$ components and suppressed population of the $|1,0\rangle$ component, where $|F,m_F\rangle$ is the hyperfine state with full magnetic moment of the atom $F$ and its projection $m_F$. The possibility of experimental realization of such a system has been demonstrated, e.g., in Ref. \cite{Leslie-et-al-2009_fluctuations}. Since the spin exchange process $(|1,1\rangle,|1,-1\rangle) \rightarrow (|1,0\rangle,|1,0\rangle)$ is suppressed, the BEC becomes effectively two-component.
For the most part of the paper, we will be interested in the 1D case when the BEC is tightly confined in the $x$ and $y$ directions by a harmonic potential; a 2D geometry will also be studied for comparison.
The two components differ only by the quantum number $m_F$, so that  equal population $N_1=N_{2}=N$ of the components results in a symmetric ground state of the system with respect to the middle-plane $z=0$.

The BEC is subject to the linear Stern-Gerlach potential produced by an external magnetic field gradient, which presses the components against each other. In the mean-field approximation the wave functions of the binary BEC obey the Gross-Pitaevskii (GP) equations:
\begin{eqnarray}
\label{eq1.01a}
i\hbar \frac{\partial }{\partial \tilde{t}}{{\tilde{\psi} }_{1}}=\left[ -\frac{{{\hbar }^{2}}}{2{{m}}}\tilde{\nabla}^2+{{\tilde{V}}_{1}}+{{g}_{11}}{{\left| {{\tilde{\psi}}_{1}} \right|}^{2}}+{{g}_{12}}{{\left| {{\tilde{\psi}}_{2}} \right|}^{2}} \right]{{\tilde{\psi} }_{1}}, \\
\label{eq1.01b}
i\hbar \frac{\partial }{\partial \tilde{t}}{{\tilde{\psi} }_{2}}=\left[ -\frac{{{\hbar }^{2}}}{2{{m}}}\tilde{\nabla}^2+{{\tilde{V}}_{2}}+{{g}_{22}}{{\left| {{\tilde{\psi}}_{2}} \right|}^{2}}+{{g}_{12}}{{\left| {{\tilde{\psi} }_{1}} \right|}^{2}} \right]{{\tilde{\psi}}_{2}},
\end{eqnarray}
where the tildes indicate dimensional variables.
The atom mass $m$ is the same for both components. The total external potential $\tilde{V}_j(\tilde{t},\tilde{z})$ provides confinement in the trap along the $x$, $y$, and $z$ axes with frequencies $\omega_{x}$, $\omega_{y}$, and $\omega_{z}$, respectively, and takes into account the magnetic Stern-Gerlach potential pressing the BEC components against each other.
In the following, for the most part only the frequency $\omega_{z}$ will be involved explicitly in our calculations. We assume tight confinement along the $y$ axis, $\omega_{y}\gg\omega_{z}$, which  produces  a Gaussian density profile in the $y$ direction \cite{bib_PethickSmith}. In the $x$ direction, we consider either tight confinement,  $\omega_{x}\gg\omega_{z}$, which leads to 1D dynamics of the system,  or $\omega_{x}=0$, which is needed to study a 2D geometry. (The exception is the 2D calculation presented in Fig.\ 15, where a finite $\omega_x$ was used.)
As a result, the potential term $\tilde{V}_j$ in Eqs. (\ref{eq1.01a}), (\ref{eq1.01b})
 reads
\begin{equation}
\label{eq1.01c}
\tilde{V}_j(\tilde{t},\tilde{z})={{m}}\omega _{z}^{2}{{\tilde{z}}^{2}}/2+{(-1)}^{j}{\mu }_{B}{B'}(t)\tilde{z}/2,
\end{equation}
where ${\mu }_{B}$ is the Bohr magneton and $j=1,2$ is the number used to label the BEC component. The Stern-Gerlach potential is turned on at the initial time instant $t=0$, ${{B'}}(t)=B'\theta(\tilde{t})$, where $B'={\rm const}$ indicates the gradient of the magnetic field magnitude, and  $\theta(\tilde{t})$ is the Heaviside step function. In the 1D case, the interaction parameters ${g_{ij}}$ are related to the respective $s$-wave scattering lengths ${a_{ij}}$ as ${{g_{ij}} \equiv 4\pi {\hbar ^2}{a_{ij}}/(\pi a_x a_y m)}$, where $a_x$ and $a_y$ are the oscillator length scales in the $x$ and $y$ (tight) directions of the trap \cite{bib_PethickSmith}, $a_{x}^2=\hbar/m\omega_{x}$, $a_{y}^2=\hbar/m\omega_{y}$. In the 2D geometry the interaction parameters are ${{g_{ij}} \equiv 4\pi {\hbar ^2}{a_{ij}}/(\pi a_y m)}$.

We assume that the inequality $g_{12}^2>g_{11}g_{22}$ is satisfied, which is the condition for phase separation.  With $g_{11}=g_{22}$ and $N_1=N_2$, the hydrostatic equilibrium for $B'=0$ implies that the densities of both BEC components are equal in the bulks, and specifically, $\tilde{n}_{01}=\tilde{n}_{02}={n}_{0}$, where $n_0$ is the peak bulk number density close to the interface in the Thomas-Fermi approximation (TFA) with neglected quantum pressure (quantum surface tension).
Initially the components 1  and 2  are located at $z<0$ and $z>0$, respectively, so that a positive $B'$ in Eqs. (\ref{eq1.01a}), (\ref{eq1.01b}) pushes the condensates against each other.

We introduce dimensionless variables as $z= \tilde{z}/a_z$, $t= \tilde{t}\omega_z/2$, and $\psi_j=\tilde{\psi_j}/\sqrt{n_0}$; then Eqs. (\ref{eq1.01a}), (\ref{eq1.01b}) read
\begin{eqnarray}
\label{1.01d}
i\frac{\partial }{\partial t}{{\psi }_{1}}=\left[ -{\nabla}^2+{{V}_{1}}+R_0^2{{\left| {{\psi }_{1}} \right|}^{2}}+R_0^2\left( 1+\gamma  \right){{\left| {{\psi }_{2}} \right|}^{2}} \right]{{\psi }_{1}}, \\
\label{1.01e}
i\frac{\partial }{\partial t}{{\psi }_{2}}=\left[ -{\nabla}^2+{{V}_{2}}+R_0^2{{\left| {{\psi }_{2}} \right|}^{2}}+R_0^2\left( 1+\gamma  \right){{\left| {{\psi }_{1}} \right|}^{2}} \right]{{\psi }_{2}},
\end{eqnarray}
where
${{V}_{j}}(t,z)={{z}^{2}}+{{\left( -1 \right)}^{j}}b(t)$.
The dimensionless GP Eqs. (\ref{1.01d}), (\ref{1.01e}) contain only three characteristic parameters of the system:
\begin{eqnarray}
\gamma\equiv \frac{g_{12}}{g}-1, \\
\label{1.02}
b=\frac{{{\mu }_{B}}{B'}{{a}_{z}}}{\hbar {{\omega }_{z}}}, \\
\label{1.03}
R_0= \sqrt{\frac{2g n_0}{\hbar {\omega_z}}}.
\label{1.04}
\end{eqnarray}
The parameter $\gamma$ measures the relative repulsion between BECs, $b$ is the dimensionless external magnetic force acting on the BECs at $t>0$, and $R_0$ is the TFA size of each condensate in the ground state at $t<0$. The TFA for each BEC component requires two conditions: (i) The system size in the $z$ direction is much larger than the respective oscillator length scale $a_z$; (ii) the interface width is much smaller than the size of the system. The parameter $R_0$ also determines the healing length in the center of the trap \cite{bib_PethickSmith}, $\tilde{\xi}_0=\hbar/\sqrt{2mgn_0}=a_z/R_0$, which is expressed in dimensionless units as
\begin{equation}
\label{eq1.05}
\xi_0=R_0^{-1}.
\end{equation}
The characteristic penetration depth of BEC density profiles may be interpreted as the interface width; it is estimated as $\tilde{\Delta}_{\rm int}=\tilde{\xi}_0/\sqrt{\gamma}$ \cite{Ao-Chui-1998,Kobyakov-2011-linear}, or, in dimensionless units
\begin{equation}
\label{eq1.06}
\Delta_{\rm int}=(\sqrt{\gamma}R_0)^{-1}.
\end{equation}
Thus, the TFA holds under conditions
\begin{equation}
\label{eq1.07}
R_0\gg1,\quad \frac{R_0}{\Delta_{\rm int}}=\sqrt{\gamma}R_0^2\gg1.
\end{equation}
The dimensionless speed of sound $c_s$ at the peak density $n_0$ is given as $c_s=\sqrt{2}R_0$.

At $t<0$ the external force $b$ is zero, and the system is in the ground state which is, in the TFA,
\begin{equation}
\label{eq1.08a}
n_{1{\rm (TF)}}(z) = |{{\psi }_{1}(z,t<0)}|^2=1-\frac{{z}^{2}}{{R}_{0}^{2}}
\end{equation}
for condensate 1 in the domain $-R_{0}<z<0$, and zero otherwise; and we find a symmetric density profile
for condensate 2 located in the region $0<z<R_{0}$.
Close to the interface, where the TFA breaks down, the interface profile can be approximated with the expression for a non-trapped system \cite{Ao-Chui-1998,bib_Waves_on_the_interface_between_two_BECs,Kobyakov-2011-linear},
\begin{equation}
\label{eq.An.Des.0.11a}
{n}_{j{\rm (int)}}(t<0,z)=\frac{1}{1+\exp\left[(-1)^{j+1} 2\sqrt{\gamma}R_0z\right]}.
\end{equation}

\section{Position swapping in a 1D trap by dynamic quantum interpenetration}
\label{sec:onedim}
We solve Eqs. (\ref{1.01d}), (\ref{1.01e}) numerically for different values of $b$ and $R_0$. In our calculations we use $\gamma=0.01$, which corresponds to the case of $^{87}\text{Rb}$ atoms in spin states $|1,\pm 1\rangle$. As an example of the numerical solution, Fig. 1 presents snapshots of the density  evolution, $n_{1,2}(t,z)$, for $b=1$ and $R_0=20$ at time instants between $t=0$ and $t=450$.
Using the standard propagation technique of the GP equations in imaginary time, we find the ground state for $t<0$ with the components separated in space as shown in Fig. 1(a). The TFA conditions, Eq. (\ref{eq1.07}), are satisfied for $R_0=20$, and the numerical solution is in agreement with
Eq. (\ref{eq1.08a}), except for the interface region close to $z=0$ and the BEC edges at $z=\pm R_0$, where quantum pressure plays an important role.
\begin{figure}
\includegraphics[width=3.55in,height=9.0in]{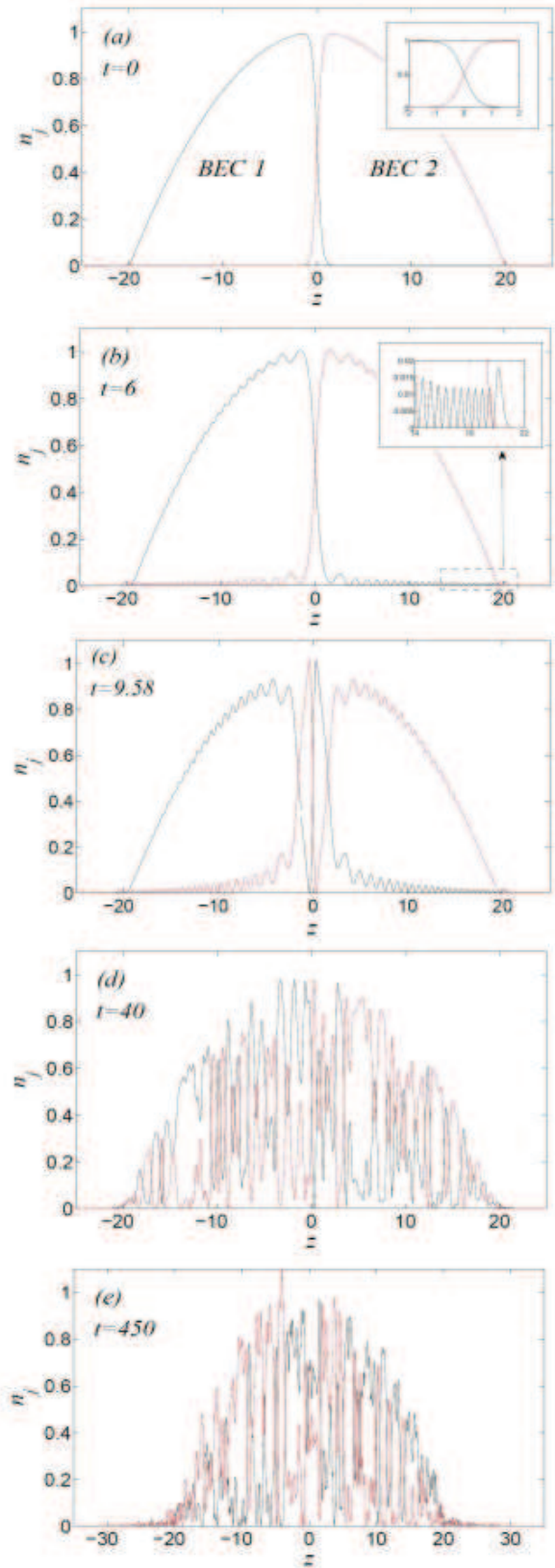}
\caption{(Color online) Snapshots of density of the BEC components at $t=0, 6, 9.58, 40, 450$ in dimensionless units.
The inset in (a) shows a magnified view of the interface. Insert in (b) presents a magnified view of the interference pattern  close to the trap edge. }
\end{figure}
\begin{figure}
\includegraphics[width=3.75in]{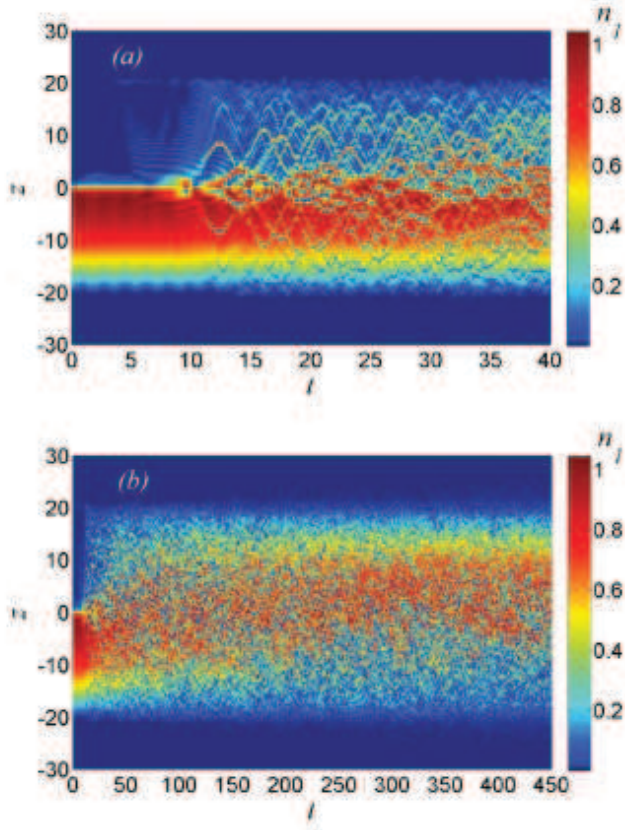}
\caption{(Color online) Evolution in time of the density $n_1$ of component 1 for $R_0=20$ in driving field $b=1$. (a) Short-time development; (b) long-time development. Component 2 is situated symmetrically with respect to $z=0$; however, at long time the detailed symmetry is spontaneously broken.}
\end{figure}
At $t=0$ we abruptly turn on the driving external potential [the second term on the right-hand side of Eq. (\ref{eq1.01c})], which presses the BEC components against  each other. For $t>0$ we solve Eqs. (\ref{1.01d}), (\ref{1.01e}) by the fourth-order Runge-Kutta method; snapshots of the evolution are shown in Fig. 1 (b)-(e). We also plot the time-resolved 
dynamics of the density $n_1(t,z)$ for component 1 in Fig.~2; the second component is situated symmetrically, \emph{i.e.} $n_2(t,z)=n_1(t,-z)$ (note, however, that at long times, numerical noise is amplified which breaks the detailed symmetry). The sudden turn-on of the driving field produces bulk oscillations of the BEC components, which may be seen in Fig. 2~(a) at $0<t<10$ at the free edge of the condensate. During the initial oscillation stage the interpenetration is minor as shown in Fig. 1~(b) and the shape of the BEC components is similar to that shown in Fig. 1~(a) with the  peak density and the condensate width oscillating in time.

The process of active quantum swapping starts with development of the first soliton of condensate 1 in the bulk of condensate 2 and vice versa at $t\approx10$, as presented in Fig. 1~(c). Eventually, more and more solitons of one BEC component penetrate the other, thus producing a random, seemingly turbulent, wave pattern. A characteristic  look of this quasi-turbulent wave pattern is shown in Fig. 1 (d) for the time instant $t = 40$, when the centers of mass of both components meet at the middle plane $z=0$. Interpenetration of the components in the form of numerous solitons may be also observed in Fig. 2. The driving force pushes the BEC components further, and they  tend to separate again in the new stable positions $z>0$ for BEC 1 and $z<0$ for BEC 2  as we can see for a relatively long time $t=450$ shown in Fig.~1~(e), and in Fig.~2~(b). Still, complete separation of the condensates is not observed even for long time intervals since the initial excess potential energy cannot disappear without energy loss processes, and it remains stored in the system in the form of kinetic energy and quantum pressure. For comparison, Fig. 3 presents the ground state of the system for the same conditions as in Fig. 1 (b)-(e), i.e., with both condensates placed in the stable positions of minimal energy from the very beginning.
\begin{figure}
\includegraphics[width=3.75in,height=2.0in]{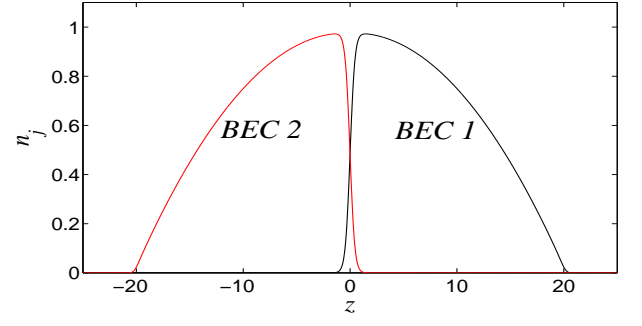}
\caption{(Color online) Ground state 
of the system for $R_0=20$ in the external potential $b=1$.
}
\end{figure}

In this paper we confine ourselves to the mean-field GP model, avoiding thermalization effects, which means that the soliton size in Fig. 1 (d)-(e) has to exceed the healing length. For this reason, we consider only moderate strengths of the magnetic field $b$. In particular, Fig. 4 presents the center-of-mass coordinates of the BEC components versus time for different values of the magnetic field $b=0.8 - 1.5$ for the system size $R_{0}=20$.  Surprisingly, even these moderate variations of the magnetic field lead to dramatic changes in the process of quantum interpenetration. In the cases $b=1$ and $b= 1.5$ the interpenetration  starts rather fast and develops actively  almost from the very beginning. Taking a slightly lower magnetic field, $b=0.9$, we observe a long preliminary stage in the system development,  for which only bulk oscillations are observed with negligible interpenetration. It requires a rather long time interval in that case, $t\approx80$, before active quantum swapping starts. Taking an even lower magnetic field, $b=0.8$, we do not observe swapping at all, which suggests the possibility of a critical magnetic field needed to drive the process.
 \begin{figure}
\includegraphics[width=3.75in,height=2.2in]{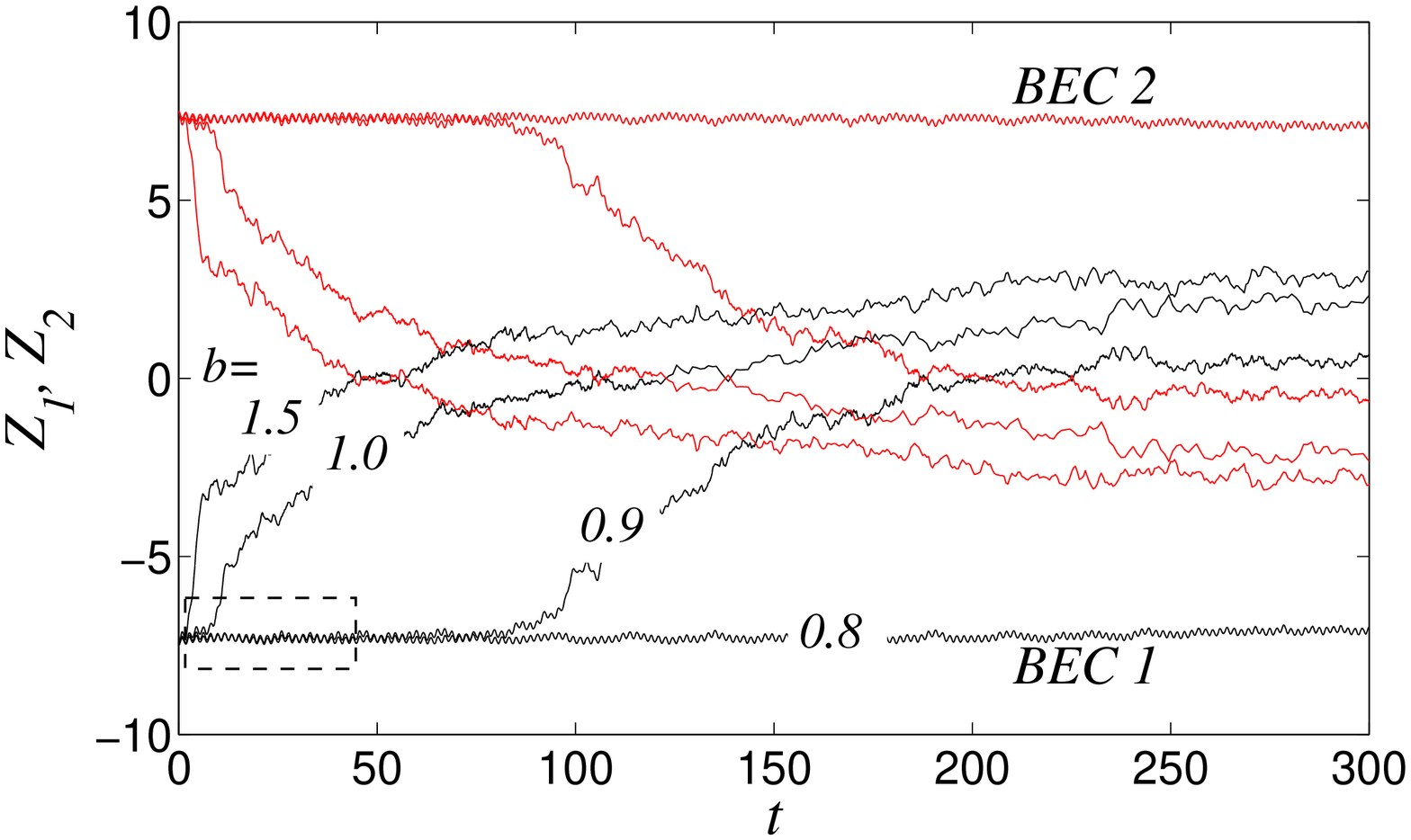}
\caption{(Color online) Positions $Z_1(t)$ and $Z_2(t)$ of center-of-mass of the BEC components for $R_0=20$ and $b=0.8, 0.9, 1, 1.5$.
}
\end{figure}

After this overview, we turn in the following subsections to discuss the different stages in the process of quantum interpenetration in detail.

\subsection{Preliminary stage: Bulk oscillations}
In this subsection we develop an analytical model for the 1D bulk oscillations of the system of two immiscible BECs and calculate the oscillation frequency. Here we assume that the driving force $b$ is sufficiently weak and does not lead to dynamic interpenetration; as we have seen, this is the case at least for short times if $b$ is small enough.
We first derive equations of motion for the center of mass (c.m.) coordinates.
Using the density-phase representation of the wave-function ${{\tilde{\psi} }_{j}}\left( t,z \right)=\sqrt{{{\tilde{n}}_{j}}}\,{\exp({i\tilde{\phi}_{j} })}$, we reduce the GP Eqs. (\ref{eq1.01a}), (\ref{eq1.01b}) to the hydrodynamic equations for the quantum fluid
\begin{eqnarray}
\label{An.Des.0.1}
&&{{\partial }_{\tilde{t}}}{{\tilde{n}}_{j}}+{{\partial }_{\tilde{z}}}\left( {{\tilde{n}}_{j}}{{\tilde{v}}_{j}} \right)=0, \\
\nonumber
&&{{\partial }_{\tilde{t}}}{{\tilde{v}}_{j}}+\frac{1}{2}{{\partial }_{\tilde{z}}}\tilde{v}_{j}^{2}=-\frac{g}{m}{{\partial }_{\tilde{z}}}{{\tilde{n}}_{j}}-\left( 1+\gamma  \right)\frac{g}{m}{{\partial }_{\tilde{z}}}{{\tilde{n}}_{3-j}} \\
\label{An.Des.0.2}
&&-\frac{1}{m}{{\partial }_{\tilde{z}}}{{\tilde{V}}_{j}}+\frac{{{\hbar }^{2}}}{2{{m}^{2}}}{{\partial }_{\tilde{z}}}\left( {\sqrt{{{\tilde{n}}_{j}}}}^{-1}{\partial _{\tilde{zz}}^{2}\sqrt{{{\tilde{n}}_{j}}}} \right).
\end{eqnarray}
where $\tilde{v}_j\equiv \hbar m^{-1} \partial_{\tilde{z}} \,\tilde{\phi}_j$ is the velocity of the quantum fluid. The boundary conditions for Eqs. (\ref{An.Des.0.1}), (\ref{An.Des.0.2}) demand that the condensates are localized,
\begin{eqnarray}
\label{eq.An.Des.0.3}
&& {\tilde{n}}_j(t,\pm\infty)=0, \\
\label{eq.An.Des.0.4}
&& \partial_{{\tilde{z}}} {\tilde{n}}_j(t,\pm\infty)=0.
\end{eqnarray}
The c.m.\ coordinate of the $j$-th component is defined as
\begin{equation}
\label{eq.An.Des.0.5}
\tilde{Z}_{j}(\tilde{t})\equiv\tilde{N}_j^{-1}\mathop{\int }^{}d\tilde{z}\,\tilde{z}\,{{\tilde{n}}_{j}(\tilde{t},\tilde{z})},
\end{equation}
where $\tilde{N}_j={\int }^{}{{\tilde{n}}_{j}}\,d\tilde{z}=n_0 a_z {\int }^{}{{n}_{j}}\,dz={\rm const}$ is the particle number.
Using Eq.(\ref{An.Des.0.1}) and integrating by parts, we obtain
\begin{equation}
\label{eq.An.Des.0.6}
{{\partial}\over{\partial {\tilde{t}}}}\left(\tilde{N}_j{{\tilde{{Z}}}_{j}}\right)=\tilde{N}_j^{-1}\mathop{\int }^{}d\tilde{z}\,{{\tilde{n}}_{j}}\tilde{V}_j.
\end{equation}
We take the second time derivative and find
\begin{eqnarray}
\label{eq.An.Des.0.7}
&&m\partial _{\tilde{t}\tilde{t}}^{2}\left(\tilde{N}_1{{\tilde{{Z}}}_{1}}\right)=-\mathop{\int }^{}d\tilde{z}\,\left( {{\tilde{n}}_{1}}{{\partial }_{\tilde{z}}}{{\tilde{V}}_{1}}+{{g}_{12}}{{\tilde{n}}_{1}}{{\partial }_{\tilde{z}}}{{\tilde{n}}_{2}} \right),\\
\label{eq.An.Des.0.8}
&&m\partial _{\tilde{t}\tilde{t}}^{2}\left(\tilde{N}_2{{\tilde{{Z}}}_{2}}\right)=-\mathop{\int }^{}d\tilde{z}\,\left( {{\tilde{n}}_{2}}{{\partial }_{\tilde{z}}}{{\tilde{V}}_{2}}+{{g}_{12}}{{\tilde{n}}_{2}}{{\partial }_{\tilde{z}}}{{\tilde{n}}_{1}} \right).
\end{eqnarray}
 In the above equations, the quantum pressure term provides zero contribution by virtue of the boundary condition, Eqs. (\ref{eq.An.Des.0.3}), (\ref{eq.An.Des.0.4}).
With the external potential $\tilde{V}_j(t,z)$ given by Eq.(\ref{eq1.01c}), and taking into  account conservation of the particle number $N_j$, we arrive at  the dimensionless form of Eqs. (\ref{eq.An.Des.0.7}), (\ref{eq.An.Des.0.8}) as
\begin{equation}
\label{eq.An.Des.0.9}
\frac{1}{4}\partial _{tt}^{2}{{Z}}_{j} +{{Z}}_{j} ={{\left( -1 \right)}^{j+1}}\left[ \frac{b}{2}+F_{12}(t) \right],
\end{equation}
where $F_{12}(t)$ is the dimensionless internal drag force between the condensates.

Thus, the c.m.\ dynamics of each component is governed by the external potentials [the first term on the right-hand-side of Eq.\ (\ref{eq.An.Des.0.9})], and by the internal resistance (drag) force of interaction between the components $F_{12}(t)$, which reads
\begin{equation}
\label{eq.An.Des.0.11}
F_{12}(t)\equiv\frac{3}{4}(1+\gamma)R_0\mathop{\int }^{}dz\,{{n}_{2}}{{\partial }_{z}}{{n}_{1}}.
\end{equation}
We stress that the derived equations are quite general, and they describe the system dynamics for all stages from bulk oscillations to active quantum interpenetration. In addition, the generalization to higher dimensions is straightforward. In particular, the equations are applicable to the steady state with no external magnetic field applied to the system, as we now show.

In the TFA in 1D geometry, the particle number and the initial c.m. coordinates are given by
\begin{eqnarray}
\label{eq.An.Des.0.5a}
&& \tilde{N}_j=n_0 a_z \frac{2R_0}{3}. \\
\label{eq.An.Des.0.5b}
&& \tilde{Z}_{0j {\rm (TF)}}\equiv \tilde{Z}_{j}(t<0)=(-1)^j a_z \frac{3R_0}{8}.
\end{eqnarray}
Eq.\ (\ref{eq.An.Des.0.9}) provides a convenient way to go beyond this approximation. In order to find the steady-state solution $\tilde{Z}_{0j} \equiv \tilde{Z}_j(t<0)$, one may calculate the interaction force $F_{12}(t<0)$ using the equilibrium profiles Eq. (\ref{eq.An.Des.0.11a}). However, it is enough to note that to first order in $\gamma$, and neglecting the curvature of the trapping potential, the components add up to form a flat interface: $n_{2{\rm(int)}}(z)=n_0-n_{1{\rm(int)}}(z)$. This immediately gives $F_{12}(t)=-(3/8)(1+\gamma)R_0n_0^2$, and as a result we obtain for the c.m. coordinates
\begin{equation}
\tilde{Z}_{0j}=(-1)^j a_z (1+\gamma)\frac{3R_0}{8}.
\end{equation}
In this way, we have found the first-order correction beyond the TFA Eq. (\ref{eq.An.Des.0.5b}).


Next, we consider an 1D solution to Eqs. (\ref{eq.An.Des.0.7}), (\ref{eq.An.Des.0.8}) in the form of small perturbations of the equilibrium state. The perturbations are produced
by the magnetic force turned on at $t=0$, assuming that the force is sufficiently weak, and the system oscillates freely at $t>0$.
The density profile of each component is represented by a bulk and an interface part,
\begin{equation}
n_1(z,t) = \left\{ \begin{array}{ll}
\tilde{n}_{1 {\rm (TF)}}(z,t), & -R_0(t) < z < -\Delta_{\rm int}, \\
\tilde{n}_{1 {\rm (int)}}(z,t), & -\Delta_{\rm int} < z < \infty, \\
0, & {\rm otherwise},
\end{array}
\right.
\end{equation}
and correspondingly for condensate 2.
Let us denote the peak density close to the trap center by $\tilde{n}_c(t)$, so that $\tilde{n}_c=n_0$ before the magnetic pulse for $t<0$. Within the TFA, we describe the unperturbed bulk profile for $t<0$ as
\begin{eqnarray}
\label{eq.An.Des.0.15}
\tnjtf^{(0)} \equiv \tnjtf(\tilde{t}<0,\tilde{z})
=n_0\left[1-\frac{\tilde{z}^2}{(a_z R_0)^2}\right].
\end{eqnarray}

We now introduce a small deviation parameter $\eta(t)$ for the central density,
\begin{equation}
\label{eq.An.Des.0.18}
\eta(t)\equiv \frac{n_c(t)}{n_0}-1.
\end{equation}
Taking into account $N_1=N_2\equiv N$, and assuming that the spatial dependence of $\tnjtf$ remains parabolic, 
we obtain the bulk density profile
\begin{equation}
\label{eq.An.Des.0.17}
\njtf\left( t>0,z \right)=
{{n}_{\rm c}}-\frac{4{{n}_{\rm c}}^{3}}{9N^{2}}{{z}^{2}}.
\end{equation}
Here we introduced the time-dependent TFA radius of the cloud, ${{R}_{TF}}( t )={3{{N}}}/[{2{{n}_{\rm c}}( t )}]$. 
Using Eq. (\ref{eq.An.Des.0.17}) we obtain $\tilde{Z}_{j}(t)=(-1)^j{9\tilde{N}}/[{16{{n}_{\rm c}(t)}}]$, and to first order in $\eta$,
\begin{equation}
\label{eq.An.Des.0.20}
\tilde{Z}_j(t>0)-\tilde{Z}_{j0}=-\eta(t)\tilde{Z}_{j0}.
\end{equation}
As above, the interface profile at $t<0$ is taken to be the infinite-system solution $\tilde{n}_{j{\rm (int)}}^{(0)}(z)$, Eq. (\ref{eq.An.Des.0.11a}), and the interface profile at $t>0$ can be found by expanding Eq. (\ref{eq.An.Des.0.11a}) to first order in the deviation $\eta$ of the central density. Again, we will be able to exploit the symmetry of the interface profiles, as we shall see shortly.
Now, neglecting the term linear in the external field $b$, Eq. (\ref{eq.An.Des.0.7}) is put on the form
\begin{equation}
\label{eq.An.Des.0.21}
m\tilde{N}\left(\partial _{\tilde{t}\tilde{t}}^{2}+\omega_z^2\right){{\tilde{{Z}}}_{1}}=-{{g}_{12}}\mathop{\int }^{}d\tilde{z}\,\left( {{\tilde{n}}_{1}}{{\partial }_{\tilde{z}}}{{\tilde{n}}_{2}} \right),\quad(t>0).\\
\end{equation}
Now insert the Ansatz, Eq. (\ref{eq.An.Des.0.20}) into the left-hand side of Eq. (\ref{eq.An.Des.0.21}). For the right-hand side, we have $F_{12}(t)=-(3/8)(1+\gamma)R_0n_0^2(1+\eta)^2$, and to first order in $\eta$ we obtain
\begin{equation}
\label{eq.An.Des.0.22}
\left[ \partial _{\tilde{t}\tilde{t}}^{2}+(3+2\gamma)\omega_z^2 \right]\eta =0.\quad(t>0)
\end{equation}
Thus, the c.m. coordinates, as well as the TFA radius and the central density, oscillate with frequency $\sqrt{3+2\gamma}\omega_z$. Let us note that the limit $\gamma\rightarrow0$ in Eq. (\ref{eq.An.Des.0.22}) cannot be taken, because when $\gamma\lesssim R_0^{-4}$, one of the assumption made for derivation of Eq. (\ref{eq.An.Des.0.22}), namely the second TFA condition in Eq. (\ref{eq1.07}), fails. In order to describe a smooth transition between the limiting case $\gamma=0$ which is equivalent to the 1-component BEC in a harmonic trap where the lowest monopole frequency is $\sqrt{2}\omega_z$, and our case restricted by Eq. (\ref{eq1.07}), one should use proper solutions for the density profiles of the ground state instead of Eq. (\ref{eq.An.Des.0.11a}).

In the dimensionless units the frequency is $2\sqrt{3.02}\approx3.48$, which may be compared with the numerical simulations:
In Fig. 5 we present the function $Z_1(t)$ found numerically for $R_0=20$ and different values of $b$, which is a magnified view of the region in Fig. 4 marked by the dashed rectangle.
\begin{figure}
\includegraphics[width=3.75in]{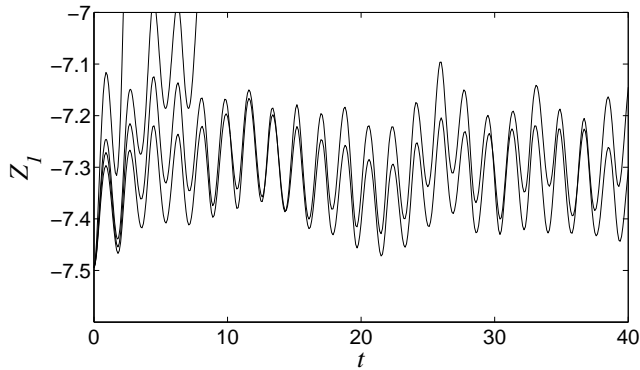}
\caption{The c.m. coordinate $Z_1(t)$ for $R_0=20$ and different values of $b$, showing the bulk oscillation.}
\end{figure}
From Fig.\ 5 we find that the oscillation frequency is $\approx3.5$, in good agreement with the analytical result. According to the numerical simulations, the frequency of bulk oscillations is almost independent of the magnetic field strength $b$. To finish this subsection, we point out that the bulk oscillations correspond to the  counter-phase mode of the BEC interface perturbations identified in the linear theory of Ref.
\cite{Kobyakov-2011-linear}, but these two modes are not identical. We remind that the counter-phase mode involves oscillations of the  interpenetration depth of the BEC components locally at the interface. For comparison, the in-phase mode corresponds to the RTI and implies bending of the interface as a whole with minor modifications of the internal structure.

\subsection{Initial stage: onset of dynamic quantum swapping}
The main purpose of the present work is to study quantum swapping of two immiscible BECs pressed against each other. Figure 1 demonstrates the  key  elements of the process, and Fig. 4 suggests that quantum swapping may occur only for a sufficiently strong driving force. Still, a certain minor interpenetration of the BEC components takes place for any magnitude of the driving force, even a low one, during the bulk oscillations. Visually, this minor initial interpenetration looks similar to that shown in Fig. 1 (b).
 As a result of this minor interpenetration, a small-amplitude matter wave of one component is injected into the ``foreign'' bulk of the other component, gets reflected from the trap edge, and forms an interference pattern. A magnified view of the interference pattern close to the trap edge is shown in the inset of Fig. 1~(b) for $b=1$, $R_0=20$. The pattern becomes even more pronounced for a stronger driving force; as an illustration, we present a snapshot  of the system evolution for $b=5$ and $R_0=20$ in Fig. 6.
\begin{figure}
\includegraphics[width=3.75in,height=2.2in]{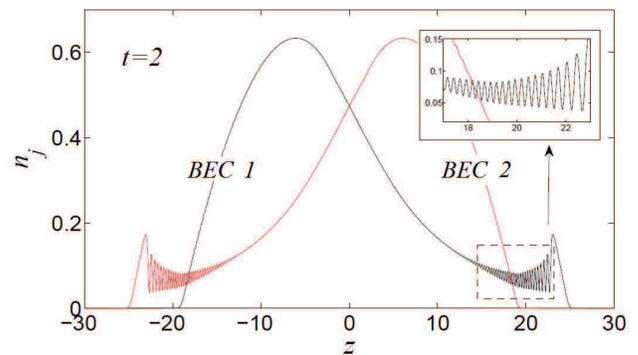}
\caption{(Color online) Snapshot of the densities at time $t=2$ for $R_0=20$ and $b=5$. The inset magnifies the interference pattern.}
\end{figure}
In the process of bulk oscillations, the amplitude of the interfering wave oscillates as well,
but these oscillations do not necessarily imply the onset of quantum swapping, since the swapping process requires unstable growth of the matter-wave amplitude.

We develop now a simplified model in order to capture qualitatively the onset of the swapping process. The model has two main ingredients: A bulk part approximated by a TFA profile, and a tail extending into the foreign component, whose growth we want to study:
\begin{equation}
n_1(z,t) = \left\{ \begin{array}{ll}
\tilde{n}_{1 {\rm (TF)}}(z,t), & -R_0(t) < z < 0, \\
\ninj(z,t), & 0 < z < R_0(t),\\
0, & {\rm otherwise}.
\end{array}
\right.
\end{equation}
Condensate 2 is given by a mirror image of the above, $n_2(z)=n_1(-z)$.
In the spirit of keeping things simple, we assume the tail to be a constant function extending across the range of the foreign component: $\tilde{n}_{1 {\rm (inj)}}(z,t) = n_0 \zeta(t)$. More elaborate expressions will not alter the qualitative results. (Note, in particular, that a spatially oscillating part should be superimposed onto the ansatz, but such oscillations will cancel out in the subsequent spatial integrations.)

The algebra carried out in Eqs.\ (\ref{An.Des.0.1})-(\ref{eq.An.Des.0.9}) can be repeated with the difference that the integration is done only over the domain $0 < z < R_0$; the resulting equation reads
\begin{eqnarray}
m\partial _{\tilde{t}\tilde{t}}^{2}\left(\int_0^{R_0}  dz\,\tilde{z}\ninj \right) =
\nonumber\\
-\int_0^{R_0}d\tilde{z}\,\ninj {{\partial }_{\tilde{z}}}\left( {{\tilde{V}}_{1}}+{{g}_{12}}{{\tilde{n}}_{2 {\rm (TF)}}} \right).
\end{eqnarray}
Note that the trap potential cancels part of the last term on the RHS by virtue of the TFA,
\begin{equation}
{\partial }_{\tilde{z}}\left(\frac12m\omega_z^2z^2 + g {\tilde{n}}_{2 {\rm (TF)}} \right) = 0,
\end{equation}
so that the equation of motion reads
\begin{eqnarray}
m\partial _{\tilde{t}\tilde{t}}^{2}\left(\int_0^{R_0}  dz\,\tilde{z}\ninj \right) =
\nonumber\\
-\int_0^{R_0}d\tilde{z}\,\ninj {{\partial }_{\tilde{z}}}\left(-\frac{\mu_B B' \tilde{z}}{2} +{({g}_{12}-g)}{{\tilde{n}}_{2 {\rm (TF)}}} \right).
\end{eqnarray}
Next, note that particle conservation implies
\begin{equation}
\label{eq.An.Des.1.1}
\tilde{n}_{\rm c}(\tilde{t})=\left[1-\frac{3}{2}\zeta(\tilde{t})\right]^{2/3}n_0.
\end{equation}
Performing the integrations and using Eq.\ (\ref{eq.An.Des.1.1}), we end up with the equation of motion for the amplitude of injected matter,
\begin{equation}
\label{eq.An.Des.1.4}
\partial _{tt}^{2}\zeta =\left[ \frac{4b}{{{R}_{0}}}-2\gamma {{\left( 1-\frac{3}{2}\zeta  \right)}^{2/3}} \right]\zeta.
\end{equation}
This equation of motion predicts that, when $b/R_0 \gg \gamma$, the injected population increases exponentially on a time scale proportional to $(2b/R_0)^{1/2}$. Moreover, it predicts a critical value of the driving force for the onset of swapping,

\begin{equation}
\label{eq.An.Des.1.5}
b_{\rm cr}=\frac{\gamma R_0}{2}.
\end{equation}
Close to the critical field, the balance of the two terms within square brackets is decisive, and the increase is super-exponential, in qualitative agreement with the cases $b=0.9, 1.0$ in Fig.\ 4.
Whether this critical field exists or whether it is an artifact of the crude approximation cannot be unambiguously inferred from the numerical results, but the behavior of the c.m. coordinate in Fig.\ 4 is suggestive.
In particular, the developed model predicts the critical force $b_{\rm cr}\approx 0.1$ for the parameters $R_0=20, \gamma=0.01$ used in the simulations of Fig. 1 - 3. This value is quite different from $b\approx 0.8$, apparent from Fig.\ 4. Still, we stress that the developed model is only qualitative and it may not be used for quantitative predictions of the system dynamics. In particular, the role of bulk oscillations were neglected in this model.

At the end of this subsection, we obtain the characteristic length scale of the interference pattern. For that purpose we evaluate the dimensionless velocity $q$ of the expanding matter wave at the trap edge from the energy conservation as
\begin{equation}
q \simeq \sqrt{2bR_0},
\label{q_standing}
\end{equation}
 where $2bR_0$ is the excess potential energy released in the expansion process. Then, the spatial dependence of the expanding matter wave close to the trap edge may be presented as $\propto \exp(\pm iqz)$, thus producing the standing wave $\propto \exp(\pm i2qz)$ with wave number $2q$ and wavelength $\lambda_{\rm sw}=\pi/q$,
\begin{equation}
\label{eq.An.Des.2.1}
\lambda_{\rm sw} = {{\pi}\over{\sqrt{2bR_0}}}.
\end{equation}
Figure 7 shows good agreement with the numerical data for the density pattern close to the trap edge at the initial stage of the quantum  interpenetration for $R_0=15,20,30$ and different values of the magnetic field.
\begin{figure}
\includegraphics[width=3.6in]{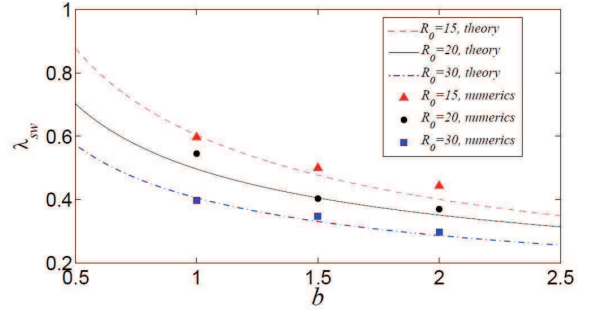}
\caption{Length scale of the standing wave at the trap edge, as function of magnitude of the driving force $b$. The curves present the analytical formula Eq. (\ref{eq.An.Des.2.1}) for $R_0=15,20,30$; the markers show the numerical result.}
\end{figure}

\subsection{Developed stage: formation of solitons and quasi-turbulent mixing of the components }
The process of active quantum swapping starts, when the first 1D "droplet" appears, i.e. bright soliton of a large amplitude $n_{1}\simeq 1$ of the component 1 develops in the bulk of the foreign component 2 (and vice versa).
Due to the pressure balance, any bright soliton of one BEC component implies a dark soliton of the other component positioned in the same place.
For driving field strength $b=1$ the first soliton  develops at $t\approx10$, as one can see in Figs. 1 (c), 2.
In the following, the soliton evolves in an essentially nonlinear way in the bulk of the foreign component, interacting with the "background" matter, and transforming into new solitons. Still, it is possible to trace  positions of individual solitons versus time during rather long time intervals in Fig. 2 (i.e. the sinusoidal "trajectories" in $z-t$ coordinates).
When the number of solitons becomes sufficiently large, the system enters the developed stage of the dynamic interpenetration process.

The developed  stage of quantum swapping exhibits complicated dynamics with large velocity fluctuations. A snapshot of the system evolution at $t=40$, presented in Fig. 1 (d), shows an example of density distribution at that stage.  At this stage, we argue that the characteristic scale of the density fluctuations must be the same as that given by the MI for two counter-flowing condensates.
The MI was analyzed in Ref.\ \cite{bib_KH_BECs} in the context of two partially overlapping, counter-flowing condensates (cf. \cite{bib_CritCounterSF}). Two initially homogeneous and coexisting condensates with relative wavevector $q$ were found to obey the dispersion relation
\begin{equation}
\label{eq.An.Des.2.2}
\frac{{{\omega }^{2}}}{2{{k}^{2}}}=\frac{{{k}^{2}}}{2}+2{{q}^{2}}+R_{0}^{2}- {{\left[ {{\left( 1+\gamma  \right)}^{2}}R_{0}^{4}+8{{q}^{2}}\left( {{k}^{2}}+R_{0}^{2} \right) \right]}^{1/2}}.
\end{equation}
\begin{figure}
\includegraphics[width=3.6in]{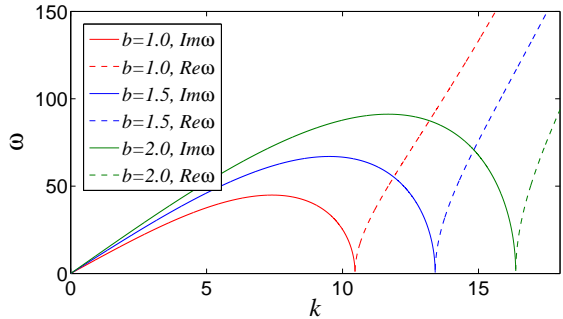}
\caption{Dispersion relation for the MI for $R_0=20$, $b=1,1.5,2$ (from bottom to top). Solid lines denote imaginary part of the frequency; dashed lines denote real part. Units are dimensionless.}
\end{figure}
\begin{figure}
\includegraphics[width=3.6in]{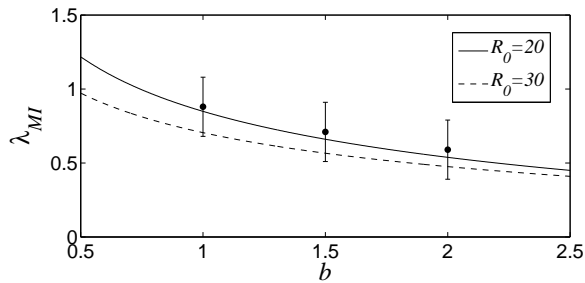}
\caption{Length scale of the multi-domain structure formed during the swapping process, as function of magnitude of the driving force $b$. The curves correspond to the analytical formula Eq. (\ref{eq.An.Des.2.4}) for $R_0=20,30$; the markers show the average numerical values obtained from the simulations for $R_0=20$; the error bars indicate the uncertainty in these values as the waves develop over a short duration of time. Units are dimensionless.}
\end{figure}
The dispersion relation Eq. (\ref{eq.An.Des.2.2}) is plotted numerically in Fig. 8 for the system size $R_0=20$ and the magnetic field strength $b=1,1.5,2$. For all parameters, we observe an unstable domain for long wavelengths, $k<k_{\rm c}$, and a stable domain at sufficiently short wavelengths, $k>k_{\rm c}$, where $k_{\rm c}$ is the MI cut-off,
\begin{equation}
\label{eq.An.Des.2.3a}
k_{\rm c}=\sqrt{2}\left[6q^{2}-R_{0}^{2}+\sqrt{(1+\gamma)^{2}R_{0}^{4}-8q^2 R_0^2+32q^{4}}\right]^{1/2}.
\end{equation}
The characteristic length scale of the instability pattern is determined by the wavelength $\lambda_{\rm MI}=2\pi/k_{\rm max}$ corresponding to the maximal growth rate. Though it is difficult to find a concise analytical formula for $k_{\rm max}$, it may be evaluated with reasonable accuracy from the numerical solution as $k_{\rm max} \approx k_{\rm c}/\sqrt{2}$, which happens to be exact at $q=0$; we obtain
\begin{equation}
\label{eq.An.Des.2.4}
\lambda_{\rm MI} = \frac{2\pi}{\left[6q^{2}-R_{0}^{2}+\sqrt{(1+\gamma)^{2}R_{0}^{4}-8q^2 R_0^2+32q^{4}}\right]^{1/2}}.
\end{equation}
We see that at $q=0$, $\lambda_{\rm MI}$ is equal within a constant factor to the interface width $\Delta_{\rm int}=(\sqrt{\gamma}R_{0})^{-1}$. The characteristic wavelength, with Eq.\ (\ref{q_standing}) inserted for $q$, is plotted in Fig. 9 together with the numerically computed wavelength, calculated as the inverse of the number of oscillations per unit length; the error bars indicate the fluctuation in this quantity over time. We note that the MI length scale is somewhat larger than the length scale of the interference pattern at the initial stage of the quantum swapping process shown in Fig. 7, albeit comparable by order of magnitude.


Eventually, the components tend to separate in space again and to occupy the positions with minimal  potential energy in the trap as shown in Fig. 1 (e) for $t=450$. In Fig. 1 (e) the main part of component 1 has moved from the domain of $z<0$ to $z>0$. This tendency may also be observed in Fig.~2~(b)
for the  density evolution of component 1, $n_1(t,z)$, presented in the space-time coordinates on large time scales. At $t=450$ shown in Fig. 1~(e), separation of the condensates is not complete; we can see remnant excitations in the condensates created  by the release of the excess potential energy and by the MI, which means that a considerable fraction of the potential energy is transformed into kinetic energy. In the next subsection we will look at energy scales in more depth.

\subsection{Rate of quantum swapping and energy balance in the process}
The overall rate of the swapping process may be characterized by the time scale, $\tau_{0}$, required for the center-of-mass of the BEC components to reach the middle of the trap, $z=0$.
The characteristic time $\tau_{0}$ is plotted in Fig.\ 10 as a function of the magnetic field strength $b$ for different sizes of the  system  $15 \leq R_{0} \leq 20$. It is seen that reducing the magnetic field $b$ or increasing the system size  $R_0$ will increase the time $\tau_{0}$ required for the swapping process. For example, by increasing the magnetic field strength from 1.0 to 2.5 for $R_{0}=20$, we decrease the swapping time by almost two orders of magnitude.
\begin{figure}
\includegraphics[width=3.6in,height=2.4in]{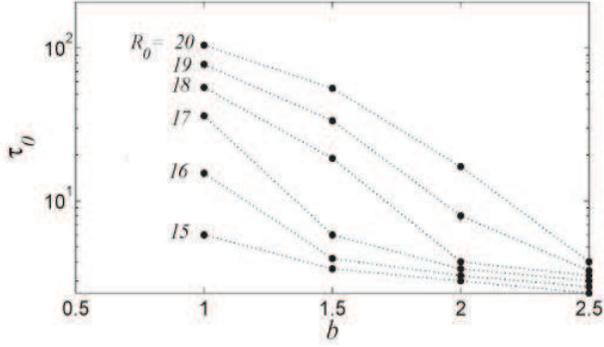}
\caption{Time $\tau_0$ after which the c.m. coordinates of the two components meet at the midpoint, as a function of the external force $b$ for different system sizes $R_0=15-20$.}
\end{figure}
Included in Fig.\ 10 are data points for those parameter sets where the swapping process is dominated by quasi-turbulent mixing. As discussed in connection with Fig.\ 4 and Eqs.\ (\ref{eq.An.Des.1.4})-(\ref{eq.An.Des.1.5}), when $R_0=20$ and $b=0.8$, no swapping was observed for the duration of the simulation, supporting our point that there exists a critical field below which there is no swapping. The case $b=0.9$, close to the critical field, appears to be governed by two quite different sorts of process; an initial slowed-down growth of injected matter in addition to the subsequent quasi-turbulent mixing.

To complete the study of the 1D quantum swapping, we discuss the balance and transformation of energies during the process. 
The total energy $E_{\rm tot}$ can be written as a sum of six contributions,
\begin{equation}
\label{eq.AnDes.1.1}
E_{\rm tot}={{E}_{\rm p}}+{{E}_{{\rm p}12}}+{{E}_{\rm t}}+{{E}_{\rm d}}+{{E}_{\rm k}}+{{E}_{\rm qp}},
\end{equation}
with the pressure energy of intra-species interactions
\begin{equation}
\label{eq.AnDes.1.6}
{{E}_{\rm p}}={\sum \limits_{j = 1,2}} \,\mathop{\int }^{}dz\,\frac{R_{0}^{2}}{2}n_{j}^{2},
\end{equation}
the pressure energy of inter-species interactions
\begin{equation}
\label{eq.AnDes.1.7}
{{E}_{\rm p12}}=\mathop{\int }^{}dz\,( 1+\gamma  )R_{0}^{2}{{n}_{1}}{{n}_{2}},
\end{equation}
the potential energy due to the trap
\begin{equation}
\label{eq.AnDes.1.5}
{{E}_{\rm t}}={\sum \limits_{j = 1,2}} \,\mathop{\int }^{}dz\,{{z}^{2}}{{n}_{j}},
\end{equation}
the potential energy due to the driving magnetic force
\begin{equation}
\label{eq.AnDes.1.4}
{{E}_{\rm d}}={\sum \limits_{j = 1,2}} \,(-1)^j\mathop{\int }^{}dz\,b z{{n}_{j}},
\end{equation}
the quasi-classical kinetic energy
\begin{equation}
\label{eq.AnDes.1.3}
{{E}_{\rm k}}={\sum \limits_{j = 1,2}}\,\mathop{\int }^{}dz\,{{n}_{j}}{{\left( {{\partial }_{z}}{{\phi }_{j}} \right)}^{2}},
\end{equation}
and the quantum pressure energy
\begin{equation}
\label{eq.AnDes.1.2}
{{E}_{\rm qp}}={\sum \limits_{j = 1,2}}\,\mathop{\int }^{}dz\,{{( {{\partial }_{z}}\sqrt{{{n}_{j}}} )}^{2}}.
\end{equation}
\begin{figure}
\includegraphics[width=3.6in]{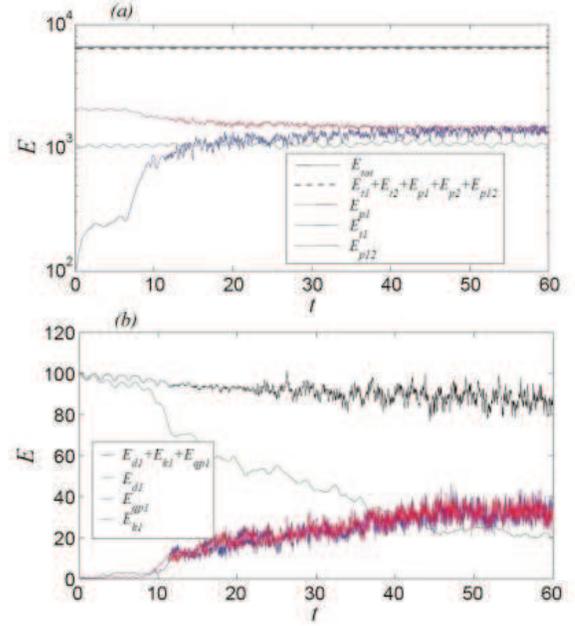}
\caption{Energy terms as functions of time. (a) Large energies. Full lines represent from top to bottom at the left, ${E}_{\rm tot}$: total energy, ${E}_{{\rm p}1}$: pressure energy of intra-species interactions for component 1, ${E}_{{\rm p}12}$: pressure energy of inter-species interactions, ${E}_{{\rm t}1}$: potential energy of component 1 due to the trap. Dashed line, almost coinciding with ${E}_{\rm tot}$, represents the sum of the large energies. (b) Small energies. Topmost line represents the sum of small energies; thereafter ${E}_{{\rm d}1}$: potential energy of component 1 due to the driving force, ${E}_{{\rm k}1}$: quasi-classical kinetic energy of component 1, ${E}_{{\rm qp}1}$: quantum pressure energy of component 1.}
\end{figure}
Since the TFA is well satisfied, and the interaction parameter $\gamma$ is small, $|\gamma|\ll1$, the energy terms in Eq. (\ref{eq.AnDes.1.1}) differ by orders of magnitude, and the whole process of energy transformation splits into two groups: The first group contains large energies related to pressure and the trap potential, and the second group  involves small energies describing  the essence of the quantum swapping.

We present the energy balance for both groups separately in Fig. 11 (a) and (b), respectively. The energies from the first group, Fig.~11~(a), are about two order of magnitude larger than the small energies from the second group, Fig. 11~(b). The sum of the energies from the first group, ${E}_{\rm p}+{E}_{\rm t}+{E}_{{\rm p}12}$, represented by the dashed line, can be hardly distinguished from the total energy, ${{E}_{\rm tot}}$, and both values are constant within a very good accuracy.

The energies from the second group provide only a minor contribution to the total energy balance, as shown in Fig.~11~(b). At the same time, the energy balance of the second group reflects the physics of the swapping process. As we can see in Fig.~11~(b), the excess potential energy in the magnetic field is released during the interpenetration, and transformed into quasi-classical kinetic energy ${{E}_{\rm k}}$ and the energy of quantum pressure ${{E}_{\rm qp}}$. The released energy is divided approximately equally between ${{E}_{\rm k}}$ and ${{E}_{\rm qp}}$. {\it A priori}, it is not obvious that small fluctuations of the large energies of the first group do not influence the energy balance in the second group. However, taking the sum of all energies of the second group, e.g. ${E}_{{\rm d}1}+{E}_{{\rm k}1}+ {E}_{{\rm qp}1}$, we find only minor variations of this value within of less than 20\% for the time span shown.


\section{Position swapping in a 2D geometry: Crossover between the RTI and quantum interpenetration}
\label{sec:twodim}
In this section we consider a 2D system, for which the position swapping can occur in two ways: either by the intrinsically multidimensional RTI or by the quantum interpenetration described above. The RTI is typically a powerful phenomenon at sufficiently large length scales, but it may be reduced and suppressed at small scales, e.g. by transport processes or quantum dispersion \cite{Cabot-Cook,Bychkov-et-al-RT-review,Bychkov-et-al-2008}.
In the quantum systems of two-component BECs, the RTI is stabilized by the effective surface tension for perturbations with wavelengths below a certain cut-off value ${\lambda}_{\rm RT}$. The cut-off was calculated in Ref.\ \cite{Kobyakov-2011-linear} using a variational Ansatz in the limit of a wide interface, and for the introduced dimensionless variables it reads as
\begin{equation}
\label{eq.2D.1}
{\lambda }_{\rm RT}=2\pi {{\left( \frac{\gamma }{4} \right)}^{1/4}}\sqrt{\frac{{{R}_{0}}}{b}}.
\end{equation}
The theoretical analysis \cite{Kobyakov-2011-linear} assumes that the inner structure of the unstable interface is not perturbed because of the 2D bending, which holds only for a driving force of limited strength. On the other hand, the quantum interpenetration studied above happens for a relatively strong magnetic field, which violates the  limits of the variational analysis \cite{Kobyakov-2011-linear}. As a result, one should expect corrections to the cut-off value,  Eq. (\ref{eq.2D.1}), due to the interpenetration effects, which make the unstable parameter domain somewhat larger. Nevertheless, we use Eq. (\ref{eq.2D.1}) as the characteristic length scale for which the RTI  becomes relatively weak.

The purpose of the present subsection is to compare the relative roles of the RTI  and quantum interpenetration at different length scales close to the analytical predictions for the RT cut-off, ${\lambda}_{\rm RT}$. In particular, we  obtain that the RTI  dominates at sufficiently large length scales $\lambda \geq \sqrt{3}{\lambda}_{\rm RT}$, corresponding to the maximal RT growth rate at the linear stage \cite{Kobyakov-2011-linear}. In that case the system dynamics resembles qualitatively the simulation results of  Refs. \cite{Sasaki-2009-RT,Kobyakov-2011-linear}, and, therefore, it is not presented here. When the system width in the transverse direction is noticeably lower than ${\lambda}_{\rm RT}$, then the RTI  does not develop at all and one should naturally expect the dominating role of dynamic interpenetration.
The respective system evolution is demonstrated for a system whose extent in the $x$ direction is $L_x=0.5{\lambda }_{\rm RT}$ in Fig. 12, which shows the density snapshots of the BEC component 1,  $n_1(x,z,t)$, at $t=0,\, 3.28,\, 12.0$, for $R_0=30$ and $b=5$. In order to simplify the analysis, here the condensate is confined by a 1D trap in the $z$ direction, and we choose periodic boundary conditions in the $x$ direction. The density of the second component 2 is not presented in Fig.~12 for brevity, but it may be reconstructed using symmetry,
which is unbroken until the quasi-turbulent regime develops. We also mark that the snapshots in Fig. 12 are strongly squeezed in the $z$ direction for illustrative purposes.
At the early stages of the system evolution presented in Fig.~12~(b), we observe a 1D pattern of stripes similar to the quantum interpenetration described in the previous subsection. Still, after a while, the 1D evolution of the system breaks down because of the multidimensional effects resembling the capillary instability \cite{capillary_instab_2011}.
\begin{figure}
\includegraphics[width=3.75in]{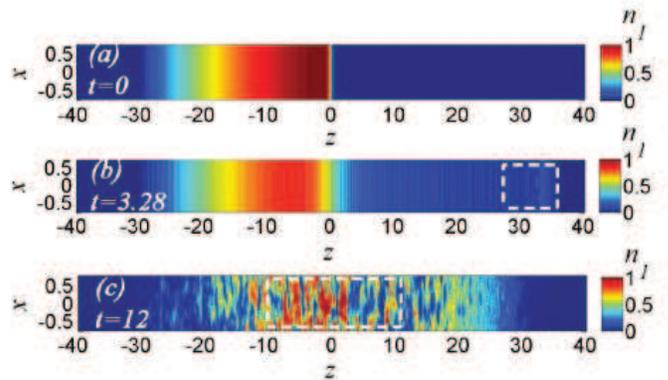}
\caption{(Color online) Snapshots of density of the BEC component $1$ for $L_x=0.5{\lambda }_{\rm RT}$, $R_0=30$ and $b=5$ at $t=0, 3.28, 12.0$ in a 2D geometry with initially flat interface between BECs. The $2$ component is located symmetrically. Note that proportions of the axes are distorted.}
\end{figure}
\begin{figure}
\includegraphics[width=3.75in]{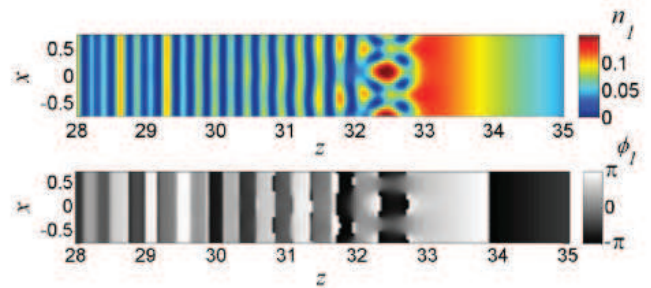}
\caption{(Color online) Magnified view of the region marked by the dashed rectangle in Fig. 13 (b): The matter-wave interference fringes break up into droplets due to the capillary instability at $t=3.28$ (note the adjusted color shading). The two panels show density $n_1$ and phase $\phi_1$ of the 1 component. The 2 component is located symmetrically.}
\end{figure}
\begin{figure}
\includegraphics[width=3.75in]{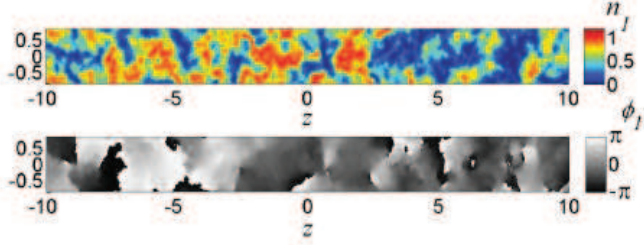}
\caption{(Color online) Magnified view of the region marked by the dashed rectangle in Fig. 13 (c): a well-developed irregular pattern of 2D droplets at $t=12.0$. The two panels show the density $n_1$ and phase $\phi_1$ of the 1 component. The 2 component is located symmetrically.}
\end{figure}
The capillary instability starts developing on the fringes, close to the system edges. In order to make the instability onset visible, Fig. 13 presents the part of the snapshot Fig. 12 (b) selected by the dashed white rectangle with equal scales in the $x$ and $z$ directions. 
The symmetry breaking develops with time into a 2D quasi-turbulent pattern of irregular droplets, as one can see  in Fig.~12 (c) for the whole trap (the snapshot is squeezed in the $z$ direction) and in Fig.~14  for the central  region with equal scales in both directions. Different from the 1D quantum interpenetration studied in the previous subsection, the multidimensional development of the MI produces 2D drops of one condensate penetrating the other instead of the 1D solitons of Fig. 1 (c)-(e). The dimensionless droplet size in Fig. 14 is about $1$, which is comparable to the soliton size in Fig. 1. 
As we have shown, the droplet size correlates with the characteristic length scale of the MI, Eq. (\ref{eq.An.Des.2.4}), which depends on the interface width between the BEC components, on the characteristic velocity of the interpenetration ``flow'' and, hence, on the magnetic field strength and the system size. This finding is also in agreement with the theoretical analysis of the quantum capillary instability performed in Ref. \cite{capillary_instab_2011}. According to the linear analysis \cite{capillary_instab_2011}, an infinitely long cylindrical rod of radius $R$ of the condensate 1 immersed in the condensate 2 breaks down into droplets by the instability with a characteristic wavelength $\lambda \simeq 4.6 R$. Further nonlinear evolution of the system in Ref. \cite{capillary_instab_2011} led to formation of droplets with the  size comparable to the initial rod diameter. Extrapolating these findings to the present study, we should expect the droplet size in our simulations to be comparable to the width of the initial 1D stripes (solitons) and, consequently, to the interface width in agreement with the present numerical data.

\begin{figure}
\includegraphics[width=3.75in]{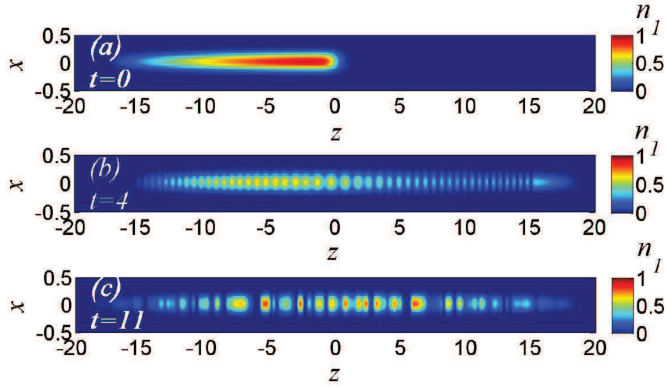}
\caption{(Color online) Snapshots of density $n_1$ of the  BEC component $1$ at $t=0, 4, 11$ in the dimensionless units of the paper in a 2D trap with transverse frequency $\omega_x=100\omega_z$ for $R_0=15$, $b=1.5$. The $2$ component is located symmetrically. Note that proportions of the axes are distorted.}
\end{figure}
On the other hand, the capillary instability cannot break the transverse symmetry when the system width is smaller than or comparable to the width of the matter-wave solitons produced by the MI. To demonstrate this we simulate the dynamics of a 2D two-component BEC trapped in both the $x$ and $z$ directions;  the simulation snapshots are shown in Fig. 15 for $R_0=15$, $b=1.5$ and $\omega_{x}/\omega_{z}=100$ at $t=0, 4, 11$. Figure 15~(a) presents the initial density for condensate 1 confined in the left half of the cigar-shaped trap, with $n_2(t,z)=n_1(t,-z)$; the snapshot (b) shows the interpenetration process at a relatively early stage, when the solitons are already formed. The solitons are strongly confined in the $x$ direction in Fig. 15, which suppresses the capillary instability. The snapshot (c) corresponds to the point when the c.m. positions of both components are close to the middle plane, $z=0$.

\begin{figure}
\includegraphics[width=3.75in]{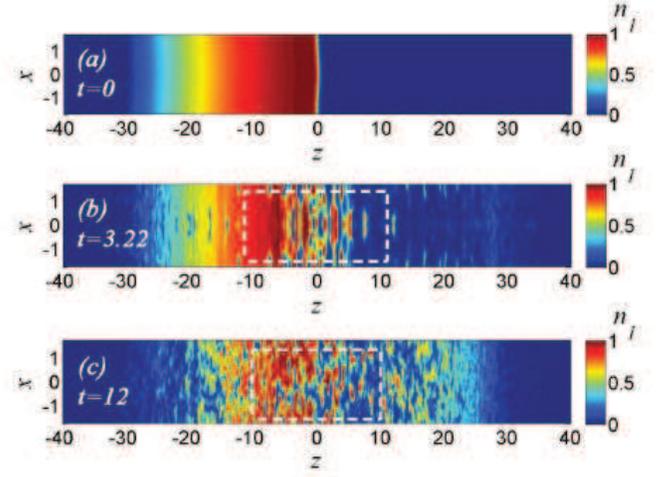}
\caption{(Color online) Snapshots of density $n_1$ of the  BEC component $1$ for $L_x={\lambda }_{\rm RT}$, $R_0=30$, $b=5$, at $t=0,3.22,12$ in a 2D geometry with initially curved interface between BECs. The $2$ component is located symmetrically. Note that proportions of the axes are distorted.}
\end{figure}
Finally, taking the system width $L_x$  close to the critical length scale ${\lambda}_{\rm RT}$ we obtain a  crossover between the RTI  and the quantum interpenetration shown in Figs. 16 for the whole system (again, the field of view is squeezed in the $z$ direction) and in Fig.~17 for the central part only (with equal scales)  for $R_0=30$ and $b=5$ at the time instants $t=0,2.84,12.0$ and $t=1.88, 1.92, 2.84, 12.0$. Though we chose the width equal to ${\lambda}_{\rm RT}$ in this simulation run, the evolution at the initial stage reproduces qualitatively the main features of the RTI, namely, unstable bending of the interface, which leads to formation of the "mushroom" structures shown in detail in Fig. 17 (a). These features of the system evolution indicate that the RT stability limits are wider at relatively strong magnetic fields, e.g. for $b=5$, in comparison with the prediction of Eq. (\ref{eq.2D.1}). We can also recognize the 2D matter-wave stripes of condensate 1 in the bulk of condensate 2 and vice versa in Fig.~16~(b) at $15<\left|z\right|<25$. However, the matter-wave stripes are rather weak as seen on the snapshot, so that they play a minor role in the system development for the width $L_x={\lambda}_{\rm RT}$, while the RTI bending dominates at the early stages. Surprisingly, further evolution of the system presented in Fig.~16~(b), (c) on large scales, and in Fig.~17~(b)-(d) in detail, differs qualitatively from the classical RT scenario.
Because of the capillary effects and quantized vortex formation, the mushroom cap detaches from the stem. One can see in Fig. 17 (b) that the vortex-antivortex pair (vortex ring in 3D) of the  component 1, which is located on the line $z\gtrsim 0$ in Fig.~17~(b) enters the spatial domain of the cap of the same component (note another vortex pair on the line $z\lesssim 0$ in Fig.~17~(b), which is just going to enter the cap). Vortex cores under the cap of the component 1 capture atoms of the component 2, and visa versa, and thus split the cap's sides into droplets, as one sees in Fig.~17~(b). The mushroom evolves into a collection of droplets, Fig.~17~(c), and eventually breaks down into a sporadic pattern, and the whole system looks more and more turbulent with time [snapshots 16 (c) and 17 (d)].
Remarkably, the sporadic system of droplets in Fig. 17 (d) demonstrates no qualitative difference from that of Fig.~14, though the latter resulted from the RTI, while the former  developed from the quantum interpenetration. This leads us to the conclusion that the quantum effects of interpenetration, surface tension and MI eventually dominate in the system dynamics, at least on the scales about ${\lambda}_{\rm RT}$, and produce quasi-turbulent quantum mixing in the system of two immiscible BECs.
\begin{figure}
\includegraphics[width=3.75in]{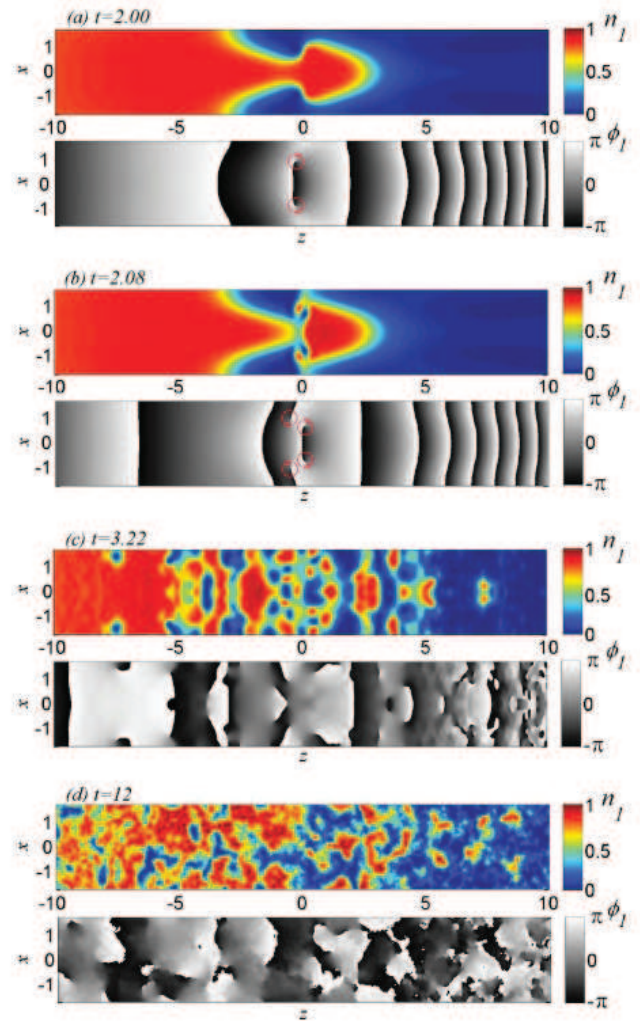}
\caption{(Color online) Snapshots of density $n_1$ of the BEC component $1$ at $t=2.00, 2.08, 3.22, 12.0$ for the same parameters as in Fig. 16. Red circular arrows in (a) and (b) show velocity direction of the BEC component $1$ around the quantized vortices.}
\end{figure}



\section{Summary and conclusions}
\label{sec:summary}
We have studied the one- and two-dimensional dynamics of a pair of immiscible Bose-Einstein condensates subject to a magnetic force acting to press the components against each other. In 1D, we find that the condensates move towards their respective potential minima through an essentially quantum-mechanical interpenetration process. The swapping process proceeds in several stages: Initial bulk oscillations accompanied by growing tails within the foreign bulk, subsequent appearance of solitons penetrating into the foreign bulk, and finally a proliferation of solitons aided by a modulational instability which results in a mixed quasi-turbulent state, in which the two condensates shift towards their new equilibrium positions. The initial part of the process is seen to happen on a short time scale, exhibiting exponential growth, except in degenerate cases. Numerical and variational calculations suggest the existence of a critical value of the driving force, below which there is no swapping.

In geometries intermediate between 1D and 2D, we study the balance between the Rayleigh-Taylor instability (RTI) and the quantum swapping, with the former dominating at the initial stage of the process provided the width of the system in the transverse direction is greater than the cut-off wavelength for the RTI. In the intermediate regime dominated by quantum swapping but not truly 1D, the interpenetration process is accompanied by capillary instabilities creating a 2D quasi-turbulent state.

\acknowledgments

This research was supported partly by the Swedish Research Council (VR) and by the Kempe Foundation. Calculations have been conducted using the resources of High Performance Computing Center North (HPC2N).


\end{document}